\documentclass[preprints,proceedingpaper,accept,moreauthors,pdftex]{Definitions/mdpi} 
\firstpage{1}
\pubvolume{9}
\issuenum{1}
\articlenumber{0}
\pubyear{2023}
\copyrightyear{2023}
\datereceived{ }
\daterevised{ } 
\dateaccepted{ }
\datepublished{ 13 December 2023}
\hreflink{https://doi.org/10.3390/psf2023009019} 

\usepackage{color,xspace}
\usepackage{dsfont}
\usepackage{subcaption}
\usepackage{natbib}
\usepackage{dsfont}
\graphicspath{{figs/}}

\setitemize{parsep=6pt,itemsep=0pt,leftmargin=*,labelsep=5.5mm}
\setenumerate{parsep=6pt,itemsep=0pt,leftmargin=*,labelsep=5.5mm} 
\setlist[description]{itemsep=0mm}   
\usepackage{environ}
\NewEnviron{myequation}{%
\begin{equation} 
\scalebox{0.95}{$\BODY$}
\end{equation}
}

\Title{Inferring Evidence 
	 from Nested Sampling Data via Information Field Theory $^\dagger$}
\TitleCitation{Inferring Evidence from Nested Sampling Data via Information Field Theory}

\Author{Margret Westerkamp
 $^{1,2}$*
, Jakob Roth $^{1,2,3}$, Philipp Frank $^{1}$, Will
 Handley $^{4,5}$ and  Torsten Enßlin $^{1,2}$} 

\AuthorNames{Margret Westerkamp, Jakob Roth, Philipp Frank, Will
 Handley, Torsten Enßlin}

\AuthorCitation{Westerkamp, M.; Roth, J.; Frank, P. ; Handley, W.; Enßlin T.}

\address{%
$^1$ \quad Max Planck Institute for Astrophysics, 85748 Garching,
 Germany
\\
$^2$ \quad Ludwig-Maximilians-Universität, Faculty for Physics, 80539
Munich, Germany
\\
$^3$ \quad Technical University Munich, Department for Computer Science,
 85748 Munich, Germany
\\
$^4$ \quad Cavendish Laboratory, Cambridge CB3 0HE, UK
\\
$^5$ \quad Kavli Institute for Cosmology, Cambridge CB3 0EZ, UK
}
\corres{Correspondence: margret@mpa-garching.mpg.de}

\firstnote{Presented at the 42nd International Workshop on Bayesian Inference and Maximum Entropy Methods in Science and Engineering, Garching, Germany, 3--7 July 2023.}

\abstract{
Nested sampling provides an estimate of the evidence of a Bayesian inference problem via probing the likelihood as a function of the enclosed prior volume. However, the lack of precise values of the enclosed prior mass of the samples introduces probing noise, which can hamper high-accuracy determinations of the evidence values as estimated from the likelihood-prior-volume function. 
We introduce an approach based on information field theory, a framework for non-parametric function reconstruction from data, that infers the likelihood-prior-volume function by exploiting its smoothness and thereby aims to improve the evidence calculation. Our method provides posterior samples of the likelihood-prior-volume function that translate into a quantification of the remaining sampling noise for the evidence estimate, or for any other quantity derived from the likelihood-prior-volume function.}

\begin{document}
\maketitle



\section{Introduction}
Nested sampling is a computational technique for Bayesian inference developed by \citep{10.1214/06-BA127}. Whereas previous statistical sampling algorithms were primarily designed to sample the posterior, the nested sampling algorithm focuses on computing the evidence by estimating how the likelihood function relates to the prior.
As discussed in \citep{Handley_2015}, Bayesian inference consists of parameter estimation and model comparison. In Bayesian parameter estimation, the model parameters $\theta_\mathcal{M}$ for a given model $\mathcal{M}$ and data $d$ are inferred via Bayes' theorem,
\begin{align}
\mathcal{P}(\theta_{\mathcal{M}}| d, \mathcal{M}) = \frac{\mathcal{P}( d| \theta_{\mathcal{M}}, \mathcal{M}) \mathcal{P}(\theta_\mathcal{M}| \mathcal{M})} {\mathcal{P}(d| \mathcal{M})} . \label{eq:BayesParameterEstimation}
\end{align}
Here, $\mathcal{P}(\theta_{\mathcal{M}}| d, \mathcal{M})$ is the posterior probability for the model parameters $\theta_\mathcal{M}$ given the data $d$. The likelihood $\mathcal{P}( d| \theta_{\mathcal{M}}, \mathcal{M})$ describes the measurement process, which generated the data $d$, and the prior $\mathcal{P}(\theta_\mathcal{M}| \mathcal{M})$ encodes our prior knowledge of the parameters within the given model. The normalization of the posterior,
\begin{align}
Z = \mathcal{P}(d|\mathcal{M}) = \int d\theta_{\mathcal{M}} \mathcal{P}(\theta_\mathcal{M}, d| \mathcal{M})\label{eq:evidence},
\end{align}
is called the evidence and is the focus of this study. 
In Bayesian parameter estimation, it is common to work with not normalized posteriors. Thus, in this scenario, the computation of the evidence is less critical. In contrast, when comparing different Bayesian models, estimating the evidence for different models is very important. In this case, the aim is to find the most probable model $\mathcal{M}_i$ given the data,
\begin{align}
\mathcal{P}(\mathcal{M}_i|d) = \frac{\mathcal{P}(d|\mathcal{M}_i)\mathcal{P}(\mathcal{M}_i)}{\mathcal{P}(d)}.\label{eq:BayesModelComparison}
\end{align}
Assuming a uniform prior for all arbitrary models
$\mathcal{P}(\mathcal{M}_i)=\text{const}$, this turns out to be equivalent to choosing the model with the highest evidence.

In nested sampling, the possibly multidimensional integral of the posterior in\linebreak  \mbox{Equation \eqref{eq:evidence}} is transformed into a one-dimensional integral by directly using the prior mass $X$. In particular, by transforming the problem into a series of nested spaces, nested sampling provides an elegant way to compute the evidence. The algorithm starts by drawing $N$ samples from the prior, called the live points. For each of these points, the likelihood is calculated and the live point with the lowest likelihood is removed from the set of live points and added to another set, called the dead points. A new live point is then sampled that has a higher likelihood value than the last added dead point. This type of sampling is commonly referred to as likelihood-restricted sampling. However, the specific methods associated with likelihood-restricted sampling are not discussed further in this paper. As a consequence of the procedure, the prior volume shrinks from one to zero, contracting around the peak of the posterior. The prior mass X contained in the parameter space volume with likelihood values larger than $L$ can be computed by
\begin{align}
\label{eq:PriorVolumeIntegral}
X(L) = \int_{\mathcal{P}( d| \theta_{\mathcal{M}}, \mathcal{M})> L} \mathcal{P}(\theta_\mathcal{M}|\mathcal{M}) ~d\theta_\mathcal{M}.
\end{align}
Thus, Equation \eqref{eq:evidence} simplifies to a one-dimensional integral,
\begin{align}
Z = \int_0^1 L(X) dX,
\end{align}
where $L(X)$ is the inverse of Equation \eqref{eq:PriorVolumeIntegral}.
Accordingly, this integral can be approximated by the weighted sum over all $m$ dead points
\begin{align}
Z \approx \sum_{i=1}^m \omega_i L_i. \label{eq:NSLevidence}
\end{align}
As proposed in \citep{10.1214/06-BA127}, we calculate the weights via $\omega_i = \frac{1}{2}(X_{i-1}-X_{i+1})$ assuming $X_0 = 1$ and $X_{m+1}=0$. Adding dead points to their set and adjusting the evidence accordingly continues until the remaining live points occupy a tiny prior volume that would contribute little to the weighted sum in Equation \eqref{eq:NSLevidence}. 

For the calculation in Equation \eqref{eq:NSLevidence} not only the known live and dead contours of the likelihood are needed but also the corresponding  prior volumes encoded in $\omega_i$, which are not precisely known. According to  \citep{10.1214/06-BA127} there are two different approaches to approximate the prior volumes $X_i$, a stochastic scheme and a deterministic scheme. In the stochastic scheme the change of volume due to each removed shell $i$ is a stochastic process characterised by a Beta distributed random variable $t_i$,
\begin{align}
X_0 =1, ~~~ X_i = t_i X_{i-1},~~~ \mathcal{P}(t_i) = \text{Beta}(t_i| 1, N ) = N t_i^{N-1} \label{eq:ChangeOfVolume},
\end{align}
where we assume a constant number of live points $N$. Approaches with a varying number of live points were i.a. introduced in dynamic nested sampling by \citep{Speagle_2020, Higson_2018} and extend beyond the boundaries of this research until the present moment.
This probabilistic description of the prior volume evolution allows to draw several samples of prior volumes $X$, according to the likelihood values $L$, and to thereby get uncertainty estimates on the evidence calculation (Equation \eqref{eq:NSLevidence}). In the deterministic scheme the logarithmic prior volume is estimated via,
\begin{align}
\ln(X_i) = -\frac{i}{N}, \label{eq:CrudeVolumeAssumption}
\end{align}
at the $i$th iteration. This estimate is derived from the fact that the expectation value of the logarithmic volume changes is $\langle \ln t_i \rangle = -1/N$. However, this estimate does not take the uncertainties in the evidence calculation \citep{Keeton_2011} into account and differs from unbiased approaches introduced and analysed in \citep{walter2015rare, Chopin_2010, salomone2018unbiased}.
In any case, the imprecise knowledge of the prior volume introduces probing noise that can potentially hinder the accurate calculation of the evidence. In order to improve the accuracy of the posterior integration, we aim to reconstruct the likelihood-prior-volume function given certain a priori assumptions on the function itself using Bayesian inference. Here, we introduce a prior and likelihood model for the reconstruction of the likelihood-prior-volume function, which we will call the reconstruction prior and the reconstruction likelihood to avoid confusion with likelihood contour and prior volume information obtained from nested sampling.

\begin{figure}[H]
  \begin{subfigure}{0.49\textwidth}
  \includegraphics[width=\linewidth]{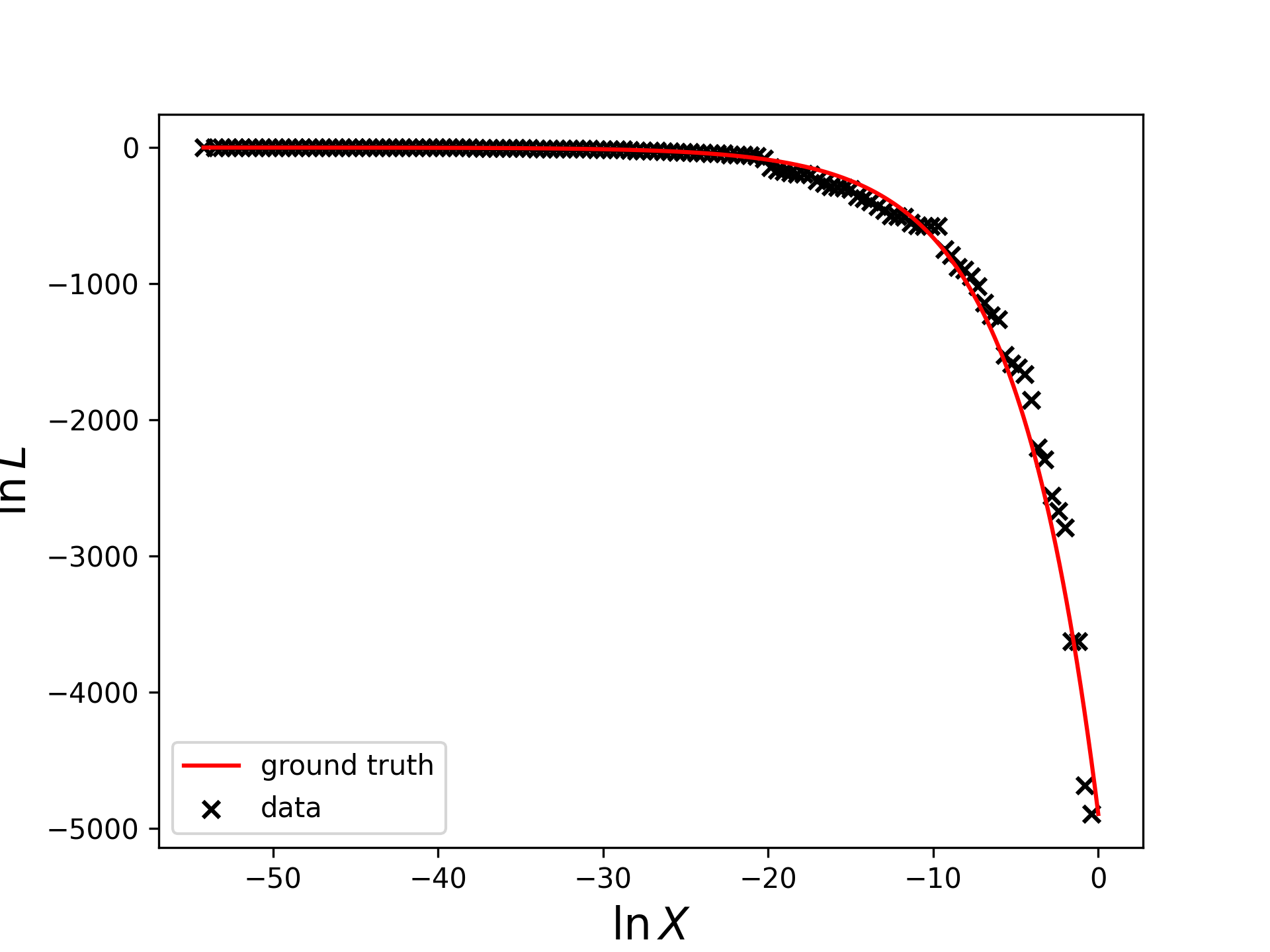}
  \end{subfigure}
  \begin{subfigure}{0.49\textwidth}
  \includegraphics[width=\linewidth]{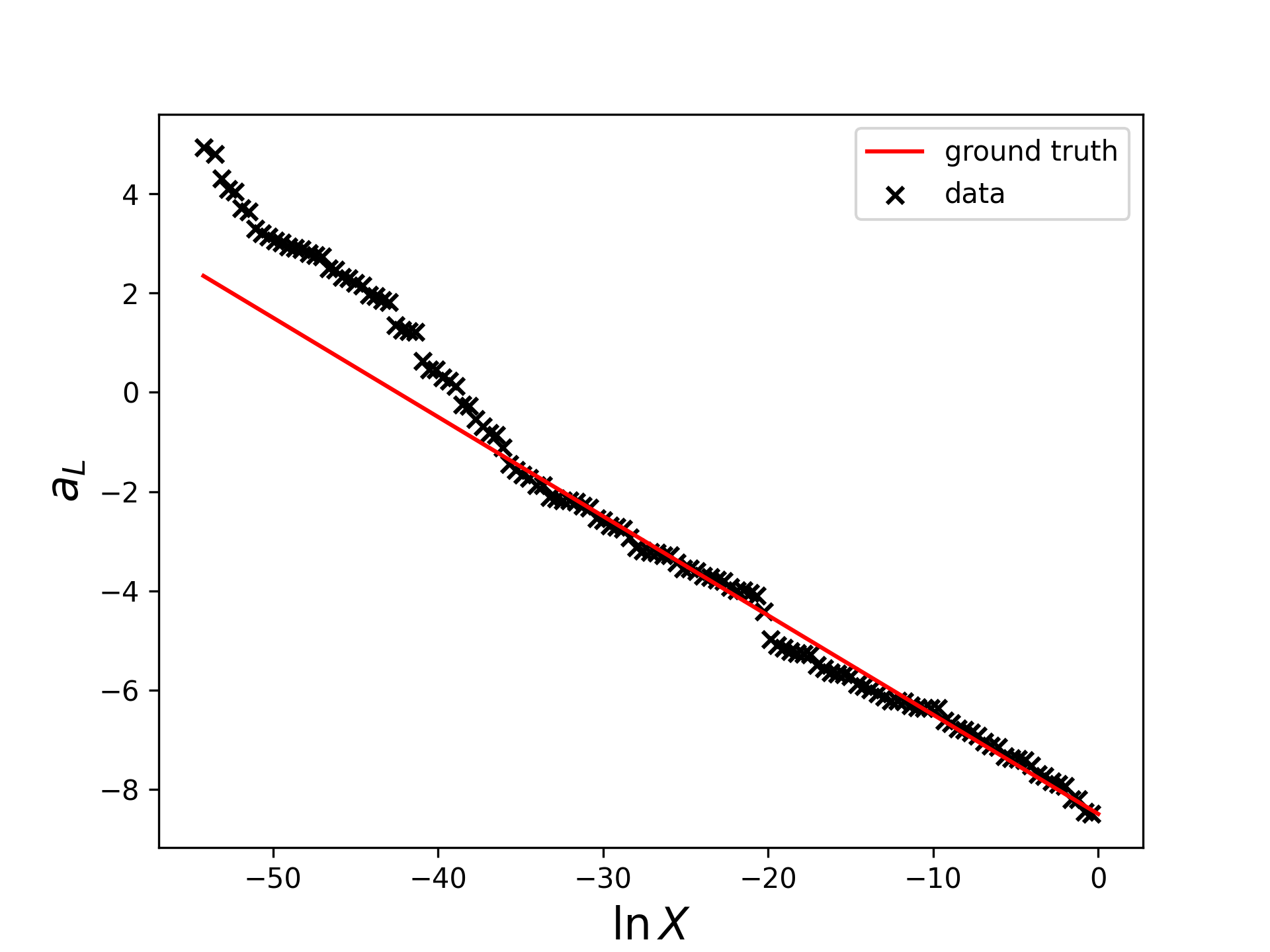}
  \end{subfigure}
  \caption{(\textbf{Left}): Visualisation of the nested sampling dead point logarithmic likelihoods, $\vec{d}_L$, as a function of logarithmic prior mass data, $\vec{d}_X$, for the normalized simple Gaussian in Equation \eqref{eq:GaussianlikelihoodCurve} ($\sigma_X = 0.01, ~D = 10$). The corresponding data was generated by the software package \mbox{anesthetic \citep{anesthetic}}. (\textbf{Right}): Visualisation of the reparametrised nested sampling dead point logarithmic likelihoods according to Equation \eqref{eq:Reparametrisation} as a function of logarithmic prior mass for the same case as shown left.}
  \label{fig:NSLdata}
\end{figure}

The left side of Figure~\ref{fig:NSLdata} illustrates the nested sampling likelihood dead contours generated by the software package anesthetic \citep{anesthetic} for the simple Gaussian example discussed in Section~\ref{sec:Parametristaion} and two live points (N = 2) as a function of prior volume. In the following, we call the likelihood values of the dead points the likelihood data $\vec{d}_L$ and the prior volume, approximated by Equation \eqref{eq:CrudeVolumeAssumption}, the prior volume data $\vec{d}_X$. Additionally, the analytical solution of the likelihood-prior-volume function, which we call the ground truth, is plotted.

\vspace{-12pt}

In accordance with the here considered example, we assume the likelihood-prior-volume function to be smooth for most real-world applications of nested sampling. In this study, we propose an approach that incorporates this assumption of a-priori-smoothness and enforces monotonicity. In particular, we use Information Field Theory (IFT) \citep{En_lin_2019} as a versatile mathematical tool to reconstruct a continuous likelihood-prior-volume function from a discrete dataset of likelihood contours and to impose the prior knowledge on \mbox{the function. }

As noted in \citep{buchner2023nested}, the time complexity of the nested sampling algorithm depends on several factors. First, the time complexity depends on the information gain of the posterior over the prior, which is equal to the shrinkage of the prior required to reach the bulk of the posterior. This is described mathematically by the Kullback-Leibler divergence (KL) \citep{10.1214/aoms/1177729694},
\begin{align}
\mathcal{D}_\text{KL} = \int d \theta_\mathcal{M} \mathcal{P}(\theta_\mathcal{M}| d, \mathcal{M}) \ln \frac{\mathcal{P}(\theta_\mathcal{M}| d, \mathcal{M})}{\mathcal{P}(\theta_\mathcal{M}|\mathcal{M})}.
\end{align} 
Second, the time complexity increases with the number of live points $N$, which defines the shrinkage per iteration. Furthermore, the time for evaluating the likelihood $L(\theta)$, $\mathcal{T}_L$, and the time for sampling a new live point in the likelihood restricted volume, $\mathcal{T}_\text{samp}$, contribute to the time complexity. Accordingly, in \citep{petrosyan2022supernest} the time complexity of the nested sampling algorithm $T$ and the error $\sigma_Z$ have been characterised via,
\begin{align}
T &\propto N \times \langle \mathcal{T}_L\rangle \times \langle
\mathcal{T}_\text{samp} \rangle \times \mathcal{D}_{\text{KL}} \label{eq:TimeComplexity} \\
\sigma_Z &\propto \sqrt{\mathcal{D}_{\text{KL}}/ N}. \label{eq:NSLError}
\end{align} 
Upon examining the error, $\sigma_Z$, it becomes evident that reducing the error by increasing the number of live points leads to significantly longer execution times.
Accordingly, by inferring the likelihood-prior-volume function, we aim to reduce the error in the log-evidence for a given $\mathcal{D}_\text{KL}$ and a fixed number of live points, $N$, avoiding a significant increase in time complexity.

The rest of the paper is structured as follows. In Section~\ref{sec:Parametristaion}, the description of the reconstruction prior of the likelihood-prior-volume curve is discussed. The model for the reconstruction likelihood and the inference of the likelihood-prior-volume function and the prior volumes using IFT is described in Section~\ref{sec:IFT}. The corresponding results for a Gaussian example and the impact on the evidence calculation are shown in Section~\ref{sec:results}. And eventually, the conclusion and outlook for future work are given in Section~\ref{sec:conclusion}.
\section{The Reconstruction Prior Model for the Likelihood-Prior-Volume Function}
\label{sec:Parametristaion}
A priori we assume that the likelihood-prior-volume function is smooth and monotonically decreasing. This is achieved by representing the negative rate of change of the logarithmic prior volume, $\ln X$, with a monotonic function of the likelihood $a_L$ as a log-normal process,
\begin{align}
-\frac{d \ln X}{d a_L} = e^{\tau(\ln X)}.\label{eq:LognormalProcess}
\end{align}
In the words of IFT, we model the one-dimensional field $\tau$, which assigns to each logarithmic prior volume a value, as a Gaussian process with $\mathcal{P}(\tau) = \mathcal{G}(\tau, T)$. Thereby, we do not assume a fixed power spectrum for the Gaussian process, but reconstruct it simultaneously with $\tau$ itself. An overview of this Gaussian process model is given in Appendix \ref{sec:CorrField}. The details can be found in \citep{Arras_2022}.

\begin{figure}[H]
  \begin{subfigure}{0.49\textwidth}
  \includegraphics[width=\linewidth]{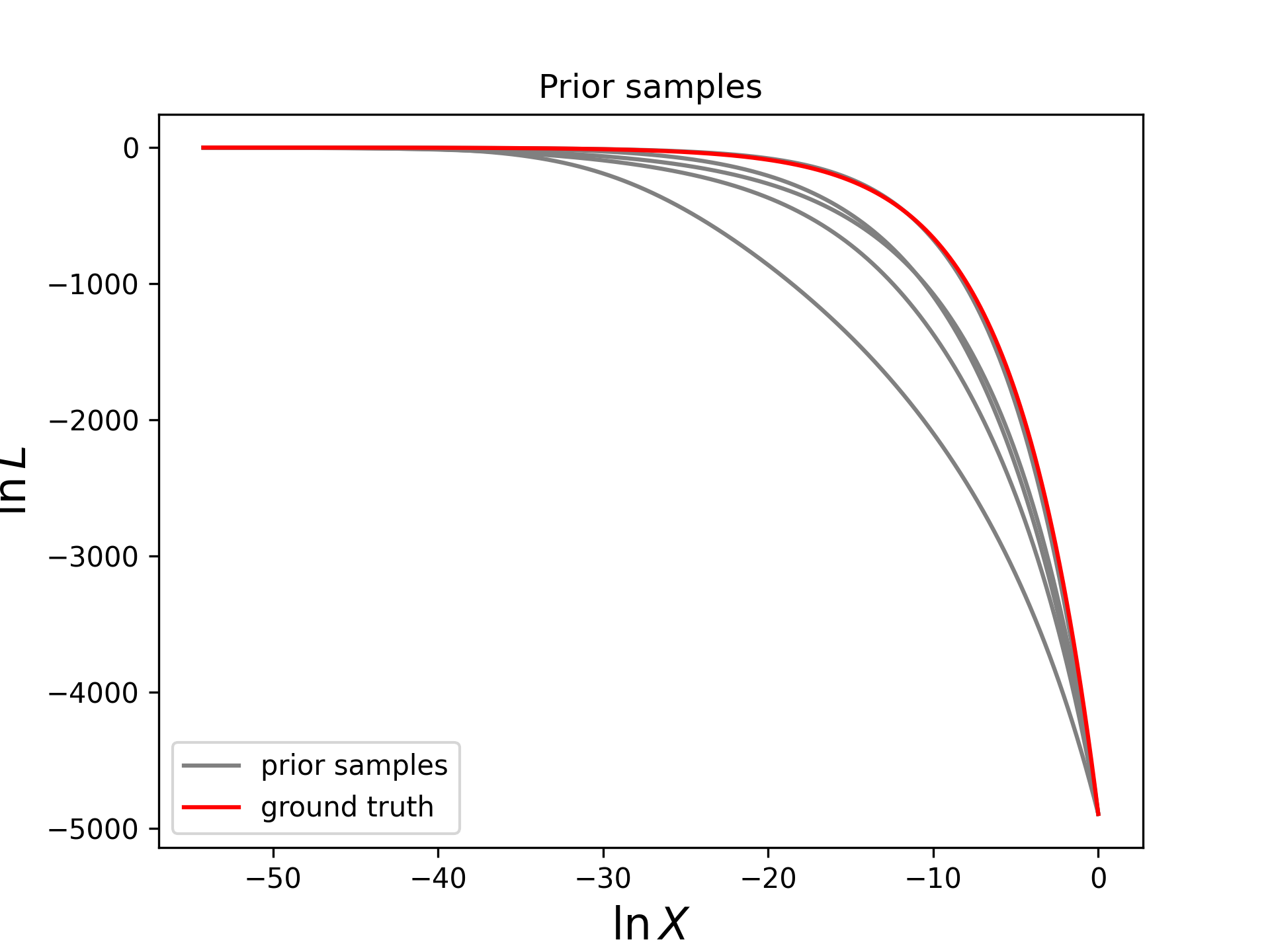}
  \end{subfigure}
  \begin{subfigure}{0.49\textwidth}
  \includegraphics[width=\linewidth]{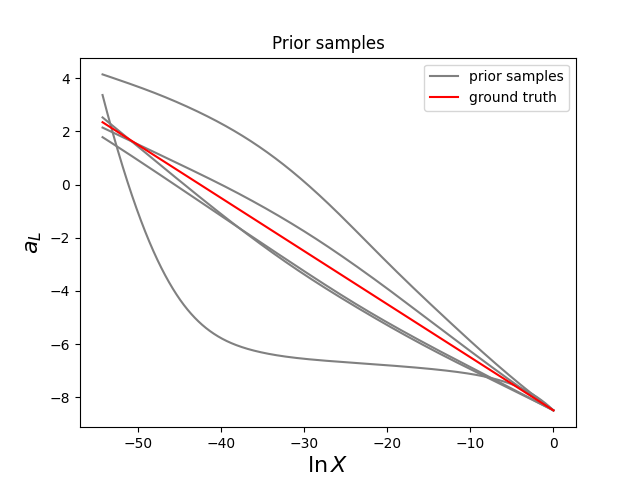}
  \end{subfigure}
  \caption{Reconstruction prior samples of the likelihood-prior-volume function plotted together with the ground truth. (\textbf{Left}): Log-log-scale. (\textbf{Right}): Parametrisation according to Equation \eqref{eq:Reparametrisation}.}
  \label{fig:PriorSamples}
\end{figure}

In the most relevant volume for the evidence, the peak region of the posterior is expected to be similar to a Gaussian in a first order approximation.  Therefore, the function $a_L$ is chosen such that $\tau$ is a constant for the Gaussian case. Deviations from the Gaussian are reflected in deviations of $\tau$ from the constant. Accordingly, we define,
\begin{align}
a_L = - \ln \biggl(-\ln \biggl(\frac{L}{L_\text{max}} \biggr)\biggr), \label{eq:Reparametrisation}
\end{align}
with $L_\text{max}$ being the maximal likelihood.
We consider the simple Gaussian example proposed by \citep{10.1214/06-BA127},
\begin{align}
L_\text{gauss}(X) &= \frac{1}{C(\sigma_X)} \exp\biggl(-\frac{X^{2/D}}{2\sigma_X^{2}}\biggr) ,\label{eq:GaussianlikelihoodCurve}
\end{align}
where $D$ is the dimension and $\sigma_X$ is the standard deviation.
We find that the function $a_L(\ln X)$, defined in Equation~\eqref{eq:Reparametrisation}, becomes linear in this case,
\begin{align}
a_\text{gauss}(\ln X) = \ln(2\sigma_X^2) - \frac{2}{D} \ln X \label{eq:GaussianLinearRelation}.
\end{align}
Figure ~\ref{fig:NSLdata} illustrates the data and the ground truth on log-log-scale on the left and the linear relation $a_L(\ln X)$ on the right. According to the log-normal process defined in \mbox{Equation \eqref{eq:LognormalProcess}}, we define the function $a_L(\ln X)$, for arbitrary likelihoods, which is able to account for deviations from the Gaussian case,
\begin{align}
a_L(\ln X) = a_0 -\int_0^{\ln X} e^{-\tau(z)} dz.  \label{eq:Functionality}
\end{align}
By inverting Equation \eqref{eq:Reparametrisation} we then get the desired likelihood-prior-volume function. The logarithmic prior volume values given the likelihood contours are obtained by inversion of Equation \eqref{eq:Functionality}.
In Figure~\ref{fig:PriorSamples} several prior samples given the model for the reconstruction prior according to \mbox{Equation \eqref{eq:Functionality}} are shown. 

\vspace{-12pt}

However, often the maximum log-likelihood, $\ln L_\text{max}$, is not known. In \citep{Handley_2019}, the calculation of the maximum Shannon entropy $\mathcal{I} = \ln \biggl( \frac{\mathcal{P}(\theta|d)}{\mathcal{P}(\theta)}\biggr)$ is given. Using this approach, we can calculate the logarithmic maximum likelihood $\ln L_\text{max}$ and thus calculate $a_L$ for unknown likelihoods.
\begin{align}
\mathcal{I}_\text{max} &= \mathcal{D}_\text{KL} + \frac{D}{2} \\
\to \ln L_\text{max} &= \langle \ln L \rangle_{\mathcal{P}(\theta_\mathcal{M}|d, \mathcal{M})} + \frac{D}{2} 
\end{align}
Hence, based on the likelihood contours obtained from the nested sampling run, we calculate the data based evidence, $Z_d$, using the approximated prior volumes according to Equation \eqref{eq:CrudeVolumeAssumption}. This allows us to obtain an estimate of the maximum log-likelihood, $\ln L_\text{max}$, of the model for reparametrisation,
\begin{align}
Z_d &\approx \sum_{i=1}^m d_{L_i} \times (d_{X_{i-1}}-d_{X_{i}}) \\
\to \ln L_\text{max} &\approx \sum_{i=1}^m \frac{d_{L_i}}{Z_d}  (d_{X_{i-1}}-d_{X_{i}}) \ln d_{L_i} + \frac{D}{2}.
\end{align}
\section{The Reconstruction Likelihood Model and Joint Inference}
\label{sec:IFT}
In this section we will derive a model for the reconstruction likelihood for the joint inference of the likelihood-prior-volume function and the changes of logarithmic prior volume according to the likelihood data $\vec{d}_L$. Here, IFT and the software package NIFTy \citep{arras2019nifty5}, which facilitates the implementation of IFT algorithms, allow us to infer the reparametrised likelihood-prior-volume function in Equation \eqref{eq:Functionality} from the data, 
\begin{align}
\vec{d}_a = - \ln \biggl(- \ln \biggl( \frac{\vec{d}_L}{L_\text{max}}\biggr) \biggr),
\end{align}
given the reconstruction prior and reconstruction likelihood model. For the inference of the likelihood-prior-volume function we first define the likelihood function $a_L$ as a function of the Beta distributed $t_i$ (Equation \eqref{eq:ChangeOfVolume}), 
\begin{align}
a_L(z_j = \sum_{i=1}^j t_i) = a_0 - \int_0^{\sum_{i=1}^j t_i} e^{-\tau(z)} dz,
\end{align}
where $a_0$ is the likelihood for $X_0 =1$. We perform a joint reconstruction of the function $a_L$ and the vector $\vec{t}_{d}$ representing changes in prior volume according to the likelihood data $\vec{d}_a$ .
\begin{align}
\mathcal{P}(a, \vec{t}_d|\vec{d}_a) &\propto \mathcal{P}(\vec{d}_a| \vec{t}_d, a) ~ \mathcal{P}(\vec{t}_d, a) \label{eq:InferenceSqueme} \\
& = \mathcal{P}(\vec{d}_a| \vec{t}_d, a) ~ \mathcal{P}(a|\vec{t}_d) ~\mathcal{P}(\vec{t}_d) \notag \\
& = \delta(\vec{d}_a - a_L(\vec{t}_d))~ \mathcal{G}(\tau|T)~ \prod_{i=1}^m \text{Beta}(t_i|1, N) \notag \\
& \approx \mathcal{G}(\vec{d}_a - a_L(\vec{t}_d), \sigma_\delta) ~ \mathcal{G}(\tau|T) ~ \prod_{i=1}^m \text{Beta}(t_i|1, N) \notag 
\end{align}
Here, the Gaussian uncertainty $\sigma_\delta$ is supposed to be chosen small in order to approximately represent the $\delta$-function. So far, we have managed to obtain a probabilistic model for the non-normalized reconstruction posterior. To this end, we use variational inference, in particular the geoVI algorithm supposed by \citep{Frank_2021}, to get an approximate of the true reconstruction posterior distribution. In the end, this statistical approach allows us to get an estimate of the likelihood-prior-volume function and the prior volumes via the posterior mean, its uncertainty and any other quantity of interest, which can be derived from the posterior samples. The results of the here developed method are shown in Section~\ref{sec:results}.
\section{Results}
\label{sec:results}
\begin{figure}[H]
\begin{subfigure}{0.49\textwidth}
  \centering
  \includegraphics[width=\linewidth]{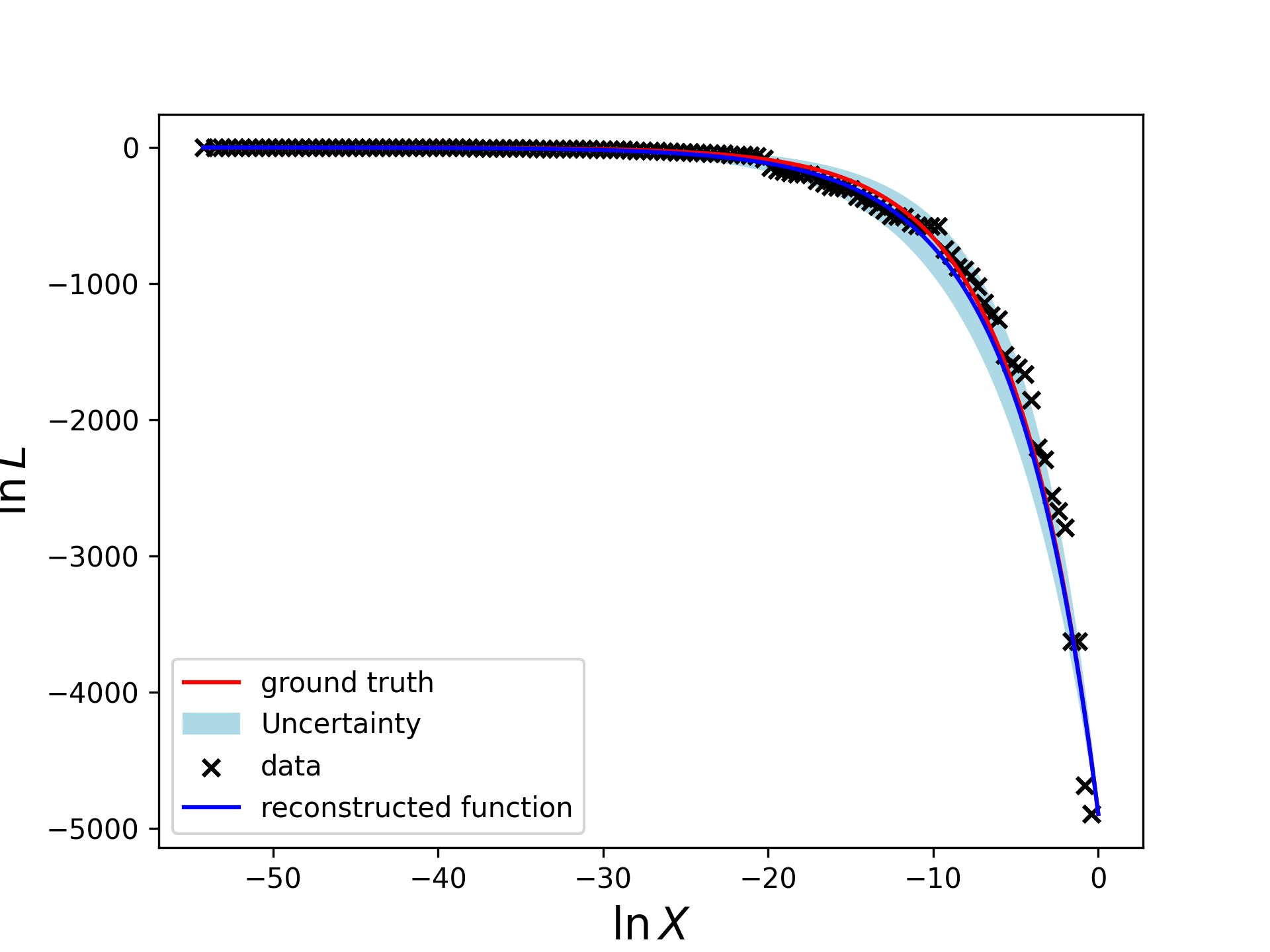}
\end{subfigure}
\begin{subfigure}{0.49\textwidth}
  \centering
  \includegraphics[width=\linewidth]{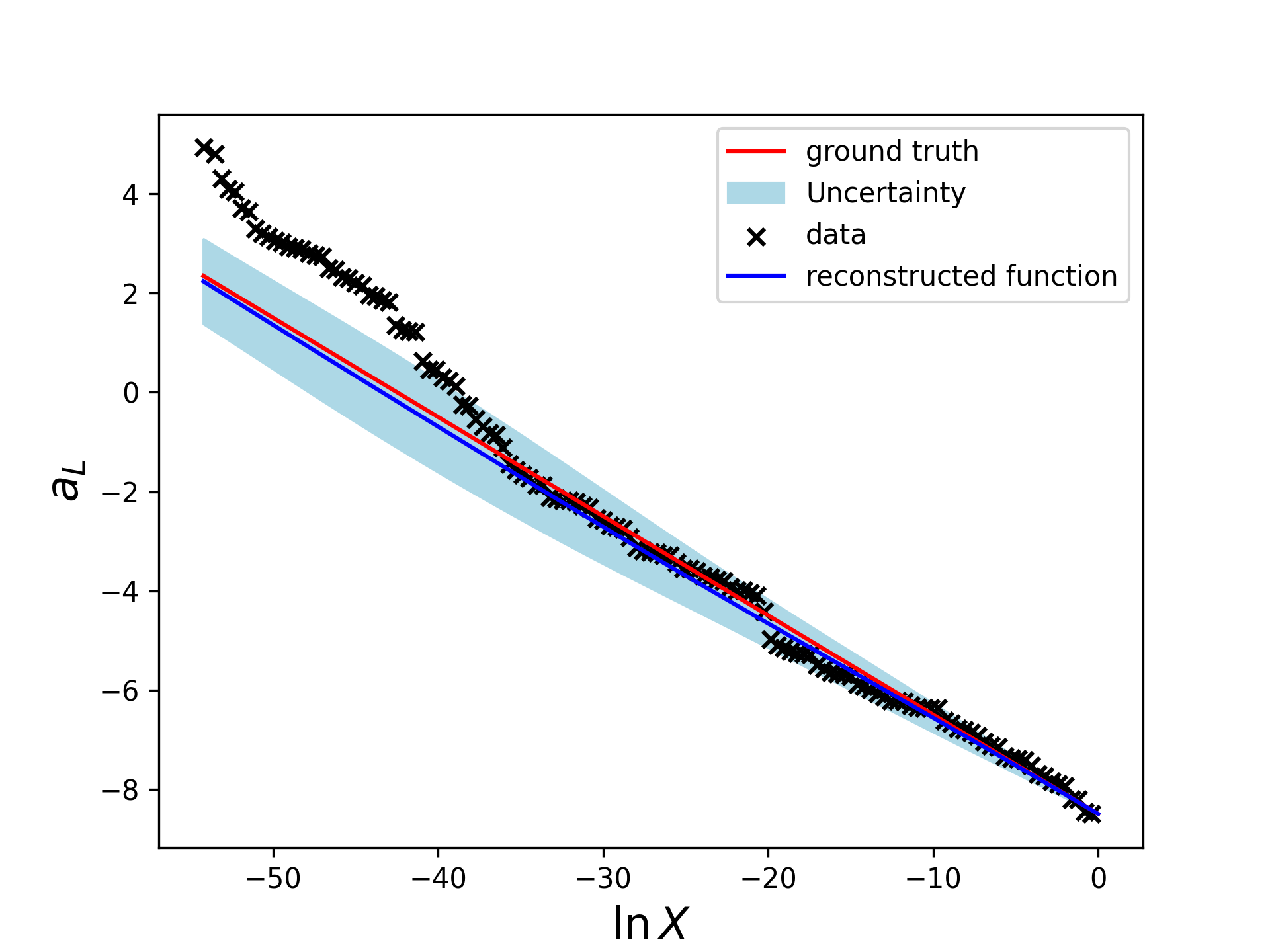}
\end{subfigure}
\caption{Reconstruction results for the likelihood-prior-volume function for the simple Gaussian example in Equation \eqref{eq:GaussianlikelihoodCurve}. The plots show the data, the ground truth and the reconstruction as well as its uncertainty. (\textbf{Left}): Reconstruction results on log-log-scale. (\textbf{Right}): Reconstruction results in reparametrised coordinates according to Equation \eqref{eq:Reparametrisation}.}
\label{fig:SimpleGaussianRec}
\end{figure}

To test the presented method we perform a reconstruction for the simple Gaussian example discussed in Section~\ref{sec:Parametristaion} and introduced in Figure \ref{fig:NSLdata}. The according results for the likelihood-prior-volume function are shown in Figure~\ref{fig:SimpleGaussianRec}. 

\vspace{-12pt}
\begin{figure}[H]
\begin{subfigure}{0.49\textwidth}
  \centering
  \includegraphics[width=\linewidth]{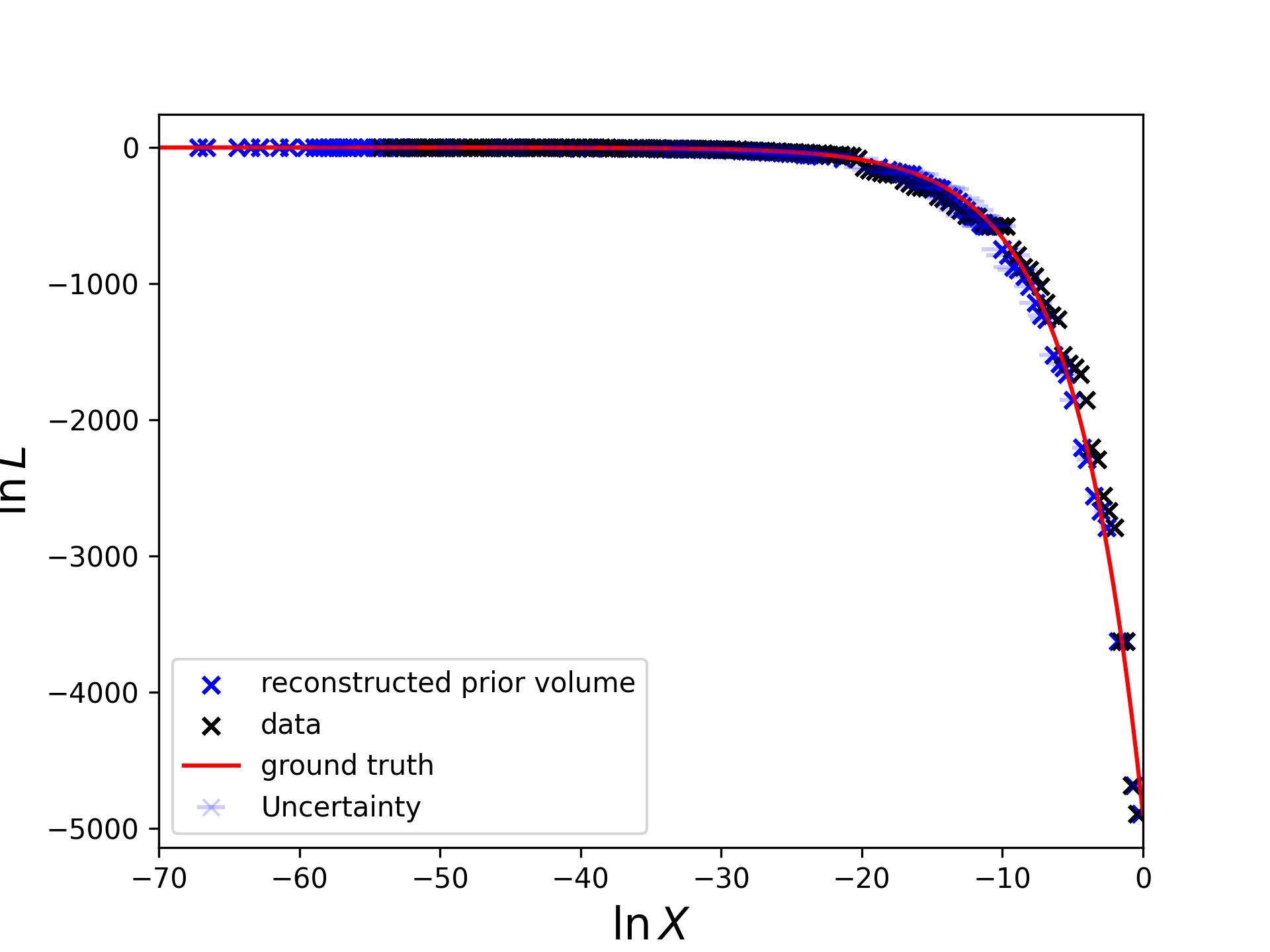}
\end{subfigure}
\begin{subfigure}{0.49\textwidth}
  \centering
  \includegraphics[width=\linewidth]{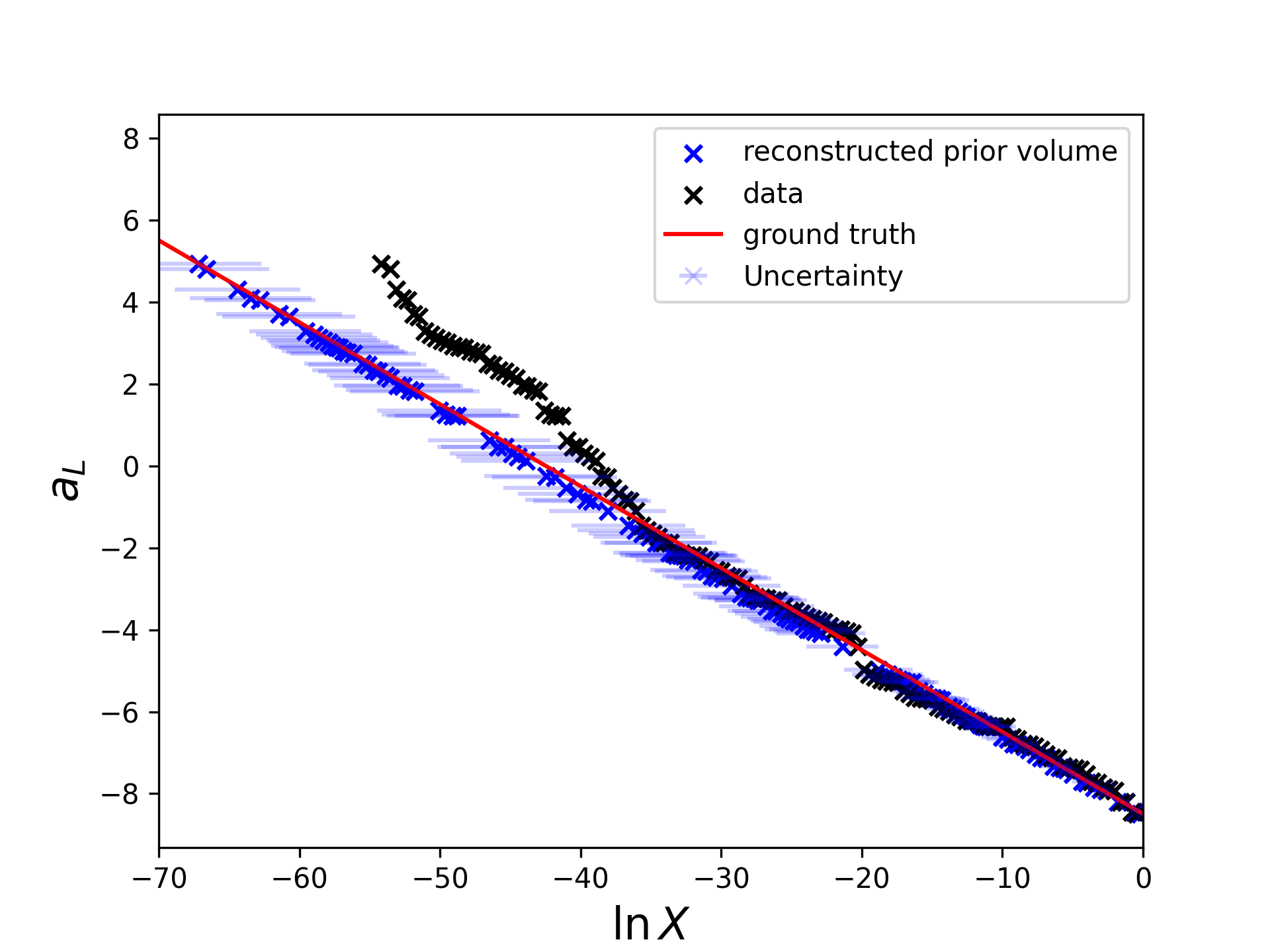}
\end{subfigure}
\caption{Reconstruction results for the prior volumes given the likelihood data $\vec{d}_L$ for the simple Gaussian example in Equation \eqref{eq:GaussianlikelihoodCurve}. The plots show the data, the ground truth and the reconstruction as well as its uncertainty. (\textbf{Left}): Reconstruction results on log-log-scale. (\textbf{Right}): Reconstruction results in reparametrised coordinates according to Equation~\eqref{eq:Reparametrisation}.}
\label{fig:SimpleGaussianRecPriorVolume}
\end{figure}

Moreover, the posterior estimates of the logarithmic prior volumes to the according likelihood data $\vec{d}_L$ are shown in Figure~\ref{fig:SimpleGaussianRecPriorVolume}.

\vspace{-12pt}
\begin{figure}[H]
  \includegraphics[width=0.6\linewidth]{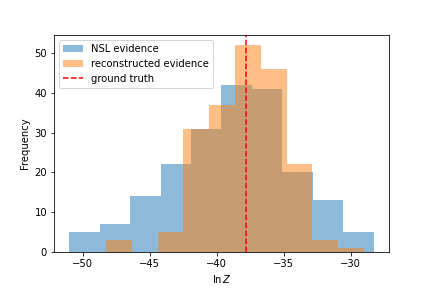}
  \caption{Comparison of histograms for logarithmic evidences for $n_\text{samp}=200$ samples for the classical nested sampling (NSL) approach and the reconstructed prior volumes.}
  \label{fig:Histogram}
\end{figure}

Since the main goal of nested sampling is to compute the evidence, we want to quantify the impact of the proposed method on the evidence calculation. To do this, we use $n_\text{samp}=200$ posterior samples for the prior volumes $X^*$ and calculate the evidence given the likelihood contours $\vec{d}_L$ for each of these samples according to Equation \eqref{eq:NSLevidence},
\begin{align}
Z^* &= \sum_{i=1}^m (\vec{d}_L)_i \frac{1}{2} (X^*_{i-1} -X^*_{i+1}). \label{eq:SampleEvidence}
\end{align}
Similarly, we generate by means of anesthetic \citep{anesthetic} $n_\text{samp}=200$ samples of the prior volume via the probabilistic nested sampling approach described in Equation \eqref{eq:ChangeOfVolume}. 
Also for these samples we calculate the evidence according to Equation \eqref{eq:SampleEvidence}. A comparison of the histograms of evidences for both sample sets (classical nested sampling and reconstructed prior volumes) is shown in Figure~\ref{fig:Histogram}.

\vspace{-15pt}

From the comparison of the histograms, one can already see that the standard deviation for the posterior sample evidences for the reconstructed prior volumes got smaller. This is also mirrored as soon as we look at numbers: The ground truth logarithmic evidence for this Gaussian case is $\ln Z_\text{gauss} = -37.798$. The result for the evidence for the classical nested sampling approach given $n_\text{samp}=200$ is $\ln Z_d = -38.92 \pm 4.50$. And finally, the result  for the evidence inferred with the here presented approach from the likelihood contours assuming smoothness and enforcing monotonicity is $\ln Z = -37.97 \pm 2.89$.
\section{Conclusions \& Outlook}
\label{sec:conclusion}
In our search for a more accurate estimate of the evidence, we set out to reconstruct the likelihood-prior-volume function. In particular, a Bayesian method was developed to infer jointly the likelihood-prior-volume function and the vector of prior volumes from the dead-point likelihood contours given by the nested sampling algorithm. For the reconstruction we enforce monotonicity and assume smoothness. 
The test of the reconstruction algorithm on a Gaussian example shows a significant improvement in the accuracy of the computed logarithmic evidence.

In general, the approach presented here will only show notable improvements if the assumption of smoothness for the likelihood- prior-volume-curve holds. Fortunately, we can reasonably expect this assumption to hold in the majority of cases. Future work will apply the reconstruction algorithm to further likelihoods where the ground truth likelihood-prior-volume function is known for testing, with the ultimate goal of applying it to actual nested sampling outputs. In particular, the results on non-Gaussian likelihoods shall \mbox{be tested.}

\vspace{6pt}
\authorcontributions{Conceptualization, M.W., J.R., P.F., W.H. and T.E.; methodology,  M.W., J.R., P.F., W.H. and T.E.; software,  M.W., J.R. and W.H.; validation, M.W., J.R., P.F., W.H. and T.E.; formal analysis, M.W., J.R., P.F., W.H. and T.E.; investigation, M.W., J.R., P.F., W.H. and T.E.; data curation, M.W., J.R. W.H.; writing---original draft preparation, M.W.; writing---review and editing, M.W., J.R., P.F., W.H. and T.E.; visualization, M.W.; supervision, T.E.; project administration, M.W.; funding acquisition, T.E.; All authors have read and agreed to the published version of the manuscript.}

\funding{Margret Westerkamp acknowledges support for this research through the project Universal Bayesian Imaging Kit (UBIK, Förderkennzeichen 50OO2103) funded
by the Deutsches Zentrum für Luft- und Raumfahrt e.V. (DLR).  Jakob Roth acknowledges financial support by the German
Federal Ministry of Education and Research (BMBF) under grant 05A20W01
(Verbundprojekt D-MeerKAT).
Philipp Frank acknowledges funding through the German Federal Ministry of Education and Research for the project ErUM-IFT: Informationsfeldtheorie für Experimente an Großforschungsanlagen (Förderkennzeichen: 05D23EO1).
This research was supported by the Munich Institute for Astro-, Particle and BioPhysics (MIAPbP) which is funded by the Deutsche Forschungsgemeinschaft (DFG, German Research Foundation) under Germany's Excellence Strategy-
EXC-2094-390783311.}

\institutionalreview{Not applicable}

\informedconsent{Not applicable}

\dataavailability{Data sharing not applicable}

\conflictsofinterest{The authors declare no conflict of interest.}

\appendixtitles{yes} 
\appendixstart
\appendix
\section{Gaussian Process Model in NIFTy\label{sec:CorrField}}

In NIFTy \citep{arras2019nifty5} we represent our reconstruction priors via generative models as described in \citep{En_lin_2022}. More precisely we use the reparametrisation trick by \citep{kingma2015variational} according to \citep{knollmueller2020metric}
 to describe the field $\tau$ with correlation structure $T$ in Equation \eqref{eq:LognormalProcess} as a generative process,
\begin{align}
\tau = A \xi, ~~\mathcal{P}(\xi) = \mathcal{G}(\xi, \Xi),~~ T = AA^\dagger .
\end{align}
Under the a priori assumption of statistical homogeneity and isotropy, $A$ becomes diagonal in Fourier space and can be fully represented by the square root of the power spectrum $p_T(|k|)$,
\begin{align}
A_{k k^\prime} &= (F A F^\dagger)_{k k^\prime} = 2\pi \delta(k - k^\prime) \sqrt{p_T(|k|)},
\end{align}
where $F$ is the corresponding Fourier transformation. Here, we model the logarithmic amplitude spectrum, $\sqrt{p_T(|k|)}$,  as a power-law with non-parametric deviations represented by an integrated Wiener process on logarithmic coordinates $l = \ln(|k|)$ according to \citep{Arras_2022}, 
\begin{align}
\sqrt{p_T(l)} \propto e^{\gamma(l)},~~ \frac{d^2 \gamma}{d l^2} = \eta \xi_\text{W} (l), ~~\mathcal{P}(\xi_W) = \mathcal{G}(\xi_W| \mathds{1}). 
\end{align}
The resulting shape of the power spectrum is encoded in $\xi_W$, $\eta$ and additional integration and normalization parameters. These additional parameters are represented by Gaussian and log-normal processes themselves and such their generative prior models are defined by hyperparameters characterising their mean and variance. 


\begin{adjustwidth}{-\extralength}{0cm}

\reftitle{References}


\PublishersNote{}
\end{adjustwidth}

\end{document}